\documentclass[conf, onecolumn]{ao4elt3}
\usepackage{graphics}
\usepackage[varg]{txfonts}
\usepackage[latin1]{inputenc}

\begin{document}
\pagestyle{plain}
\title{LOCAL ENSEMBLE TRANSFORM KALMAN FILTER: A NON-STATIONARY CONTROL LAW FOR COMPLEX ADAPTIVE OPTICS SYSTEMS ON ELTS}
\author{GRAY Morgan\inst{1}\thanks{morgan.gray@lam.fr} \and PETIT Cyril\inst{2} \and RODIONOV Sergey\inst{1} 
\and  BERTINO Laurent\inst{3} \and BOCQUET Marc\inst{4,5} \and FUSCO Thierry\inst{1,2}}
\institute{Aix Marseille Universit\'e, CNRS, LAM (Laboratoire d'Astrophysique de Marseille) UMR 7326, 13388 Marseille, France  \and ONERA,  29 avenue de la Division Leclerc,  92322 Ch\^atillon, France
\and NERSC,  Thormohlens gate 47,  N-5006 Bergen, Norway \and Universit{\'e}  Paris-Est, CEREA joint laboratory \'Ecole des Ponts ParisTech and EDF R\&D, 6-8 avenue Blaise Pascal,  77455 Marne la Vall\'ee, France \and INRIA, Paris Rocquencourt research center, France}


\abstract{
We propose a new algorithm for an adaptive optics system control law which allows to reduce the computational burden in the case of an Extremely Large Telescope (ELT) and to deal with non-stationary behaviors of the turbulence. This approach, using Ensemble Transform Kalman Filter and localizations by domain decomposition is called the local ETKF: the pupil of the telescope is split up into various local domains and calculations for the update estimate of the turbulent phase on each domain are performed independently. This data assimilation scheme enables parallel computation of markedly less data during this update step. This adapts the Kalman Filter to large scale systems with a non-stationary turbulence model when the explicit storage and manipulation of extremely large covariance matrices are impossible. First simulation results are given in order to assess the theoretical analysis and to demonstrate the potentiality of this new control law for complex adaptive optics systems on ELTs.}
\maketitle


\section{Introduction}\label{intro}

In order to reach optimality (i.e. minimization of the residual phase variance), control laws for new Adaptive Optics (AO) concepts in astronomy require the implementation of techniques intended for the real time estimation of the atmospheric turbulence [1]. It has been proven that the Kalman Filter (KF) based optimal control law enables the estimation and prediction of the turbulent phase and efficiently corrects the various modes of this phase in the case of tomographic or wide field AO systems (LTAO, MOAO and MCAO) [2-4]. However, for any ELT with a basic implementation of a KF-based control law, the numerical complexity of all computations involved in matrix calculations will be an insuperable burden so that it will be impossible to deal with non-stationary behaviors. We have previously proposed a new control law based on the Ensemble Transform Kalman Filter (ETKF) [5,6], which offers a non-stationary model of the turbulence and a parallel environment implementation. This control law can reach the performances of the KF-based control law: but at the cost of dramatically increasing the number of members in the ensemble. Here, we propose a new version, the local ETKF, which allows to significantly reduce this computational cost in the case of an ELT with a natural parallel implementation and to handle a non-stationarity model of the turbulence.\\This article is organized as follows: section 2 recalls the main ideas of the KF and the ETKF-based control laws, gives the limitations of the ETKF and explains why the local ETKF is highly suitable to overcome all drawbacks. Section 3 presents two numerical simulations: convergence of the ETKF performances to the KF performances, the better results obtained with the local ETKF for different numbers of members and 2 numerical examples of theoretical complexity. Section 4 briefly discusses parallelization and preliminary speed tests with our first implementation of the local ETKF. Some conclusions and an account on the work in progress are given in section 5.


\section{The local ETKF: a suitable control law for AO on ELTs}\label{sec:1}

In a previous paper [6], we have already described in detail the mathematical structure of the KF and of the ETKF-based control laws. Let us recall the main concepts.

\subsection{From the KF...}

The Kalman Filter (KF) is based on a stochastic linear state-space model which can be defined with two equations. The first equation characterizes the dynamics of the turbulent phase with an Auto Regressive model (AR 1 or AR 2) and a vector $x_{k}$ containing two successive instants of the turbulent phase. With a zonal basis, the phase is estimated on the locations of the Deformable Mirror's (DM) actuators: therefore, the dimension of the vector $x_{k}$ is \textit{n}, where $n = 2\, \times$ number of actuators [6]. The second equation is the observation equation with a vector $y_{k}$ containing the slopes of the residual phase (in two directions) on each subaperture given by the Shack-Hartmann Wave Front Sensor (S-H WFS): thus, the dimension of the vector $y_{k}$ is \textit{p}, where  $p = 2\, \times$ number of sub pupils [6]. In the linear Gaussian case, the optimal solution of the prediction estimate of $x_{k}$ is:
\begin{equation}
\hat{x}_{k+1/k} \,=\, {A}_{1}\hat{x}_{k/k-1} \,+\, {A}_{1}H_{k}[y_{k} \,-\, \hat{y}_{k/k-1}]
\end{equation}
where $H_{k}$ is the Kalman gain. With a basic KF implementation, we must solve the Riccati matrix equation [1,6] in order to calculate this Kalman gain $H_{k}$. If the model is stationary, it can be done off-line by taking the asymptotic solution: however, as the theoretical numerical complexity is $O(n^{3} + p^{3})$, this standard solver is very slow for AO systems on ELTs. In order to use KF on ELT, solutions have been proposed [7], but accounting for the non-stationary behavior of the turbulence during observation time with a reduced complexity is still a hard point.

\subsection{...through the ETKF...}

The Ensemble Transform Kalman Filter (ETKF) has been developed in geophysics in order to overcome the previous limitations (stationarity, numerical complexity for transition to a very large number of parameters). The two main ideas of the ETKF [8,9] are based on:\\
- firstly, the creation of an initial ensemble of \textit{m} members. This ensemble yields update and prediction estimates of the turbulent phase on the pupil of the telescope,\\
- secondly, the substitution of the real covariance matrices by the empirical covariance matrices calculated from the \textit{m} members estimates of this ensemble.\\
With the ETKF-based control law, the original equation (1) with the Kalman gain $H_{k}$, containing the previous empirical covariance matrices, is totally transformed [8-10] in order to reduce the numerical complexity of the computations in the case of a non-stationary turbulence model. The theoretical numerical complexity of the ETKF is $O(m^{3} + m^{2}\times(n + p))$ [10]: therefore, the number \textit{m} of members in the ensemble must remain much smaller than the values of n and p.\\
However, to obtain the same results as those given by the KF optimal solution, it is necessary to dramatically increase this number \textit{m} of members in the ensemble (see section 3.1).\\
Similarly, when the diameter of the telescope goes from 8 m to 40 m, it is necessary to dramatically increase again this number of members. The ETKF is computed in a space spanned by the \textit{m} members of the ensemble: this subspace is rather small compared to the total dimension of the model space on the whole pupil surface of the telescope (it will be especially true for an ELT: see numerical example in section 3.2). Thus, the empirical covariance matrices calculated from the ensemble will not be able to match the model rank (the number of degrees of freedom of the model) [11].\\
Moreover, with the ETKF-based control law, we use a finite ensemble size to approximate the true estimation error covariance matrices: it will introduce sampling errors that are seen as spurious correlations [12-14] over long spatial distances on the whole pupil of the telescope. Some variables known to be uncorrelated with an observation, will introduce some small unphysical updates. Therefore, over time, with each spurious update, the empirical covariance matrices calculated from the ensemble are underestimating the true covariance matrices.

\subsection{...to the local ETKF}

In order to solve the previous problem of subspace dimension much lower than the one of the whole pupil, a new version called the local ETKF [14] has been developed using domain decomposition and localizations. The pupil of the telescope is split up into various local domains in which all calculations of each update step are performed independently [6].\\
Figure 1 shows the locations of the actuators (blue dots where the phase is estimated) on the pupil of a 16 m diameter telescope sampled by a 32$\times$32 SH-WFS with a Fried geometry (the square areas between 4 dots represent the subapertures).
\begin{figure}[!ht]
\centerline{\resizebox{0.45\columnwidth}{!}{\includegraphics{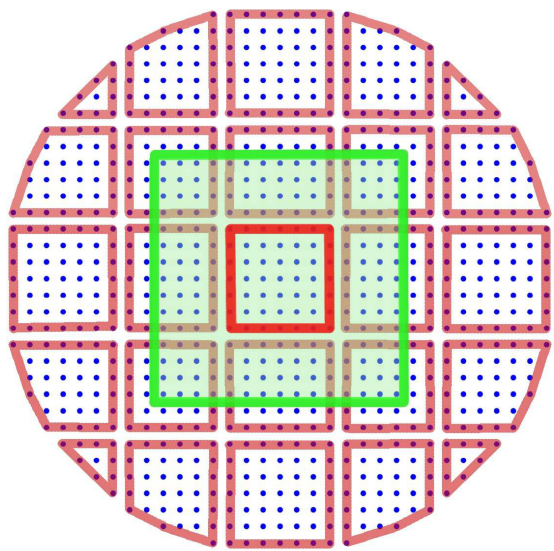}}}
\caption{Partition of the actuators domain of a 16 m diameter telescope's pupil (with a 32$\times$32 SH-WFS)}
\label{fig:1}
\end{figure}
\\
In this basic example, the domain decomposition is made of 25 red estimation subpartitions (so-called estimation domain), each of them composed by a fixed number of actuators.\\
During the update step, the update estimate calculated from the slopes given by the SH-WFS [6] (so-called data assimilation) is performed locally with an observation region made up of subapertures surrounding each estimation domain: for instance, the data assimilation region for the central estimation domain is the green one. This scheme enables parallel computation of markedly less data during each update step.\\
Therefore, this local ETKF-based new control law allows the realization of a parallel environment implementation with Central Processing Units (CPUs) or Graphics Processing Units (GPUs): see short discussion in section 4.


\section{Zonal Single Conjugate AO simulations with the ETKF \& the local ETKF}\label{sec:2}

In order to validate some of the items discussed in section 2, we have implemented the three previous control laws under Matlab environment   and made comparisons in term of coherent energy. For the simulation of the atmosphere, we consider a Von Karman turbulence [7]: $r_{0}$ \,=\, 0.525 m, $L_{0}$ \,=\, 25 m, $\lambda$ \,=\, 1.654 $\mu$m (for both WFS's and observation's wavelengths). Using Taylor's hypothesis, we can generate a superimposition of 3 turbulent phase screen layers moving at 7.5 $ms^{-1}$, 12.5 $ms^{-1}$, 15 $ms^{-1}$, with a relative $C_{n}^{2}$ profile of 0.5, 0.17, 0.33 respectively.\\
For this zonal Single Conjugate AO (SCAO) system simulation with three different control laws (KF, ETKF and local ETKF), we consider 2 cases: a 8 m telescope (section 3.1)  and 16 m telescope (section 3.2) with respectively a 16$\times$16 and a 32$\times$32 microlenses S-H WFS with a transmission factor equal to 0.5: the linear response of the S-H WFS is emulated by a matrix D which calculates difference of the phase at the edges of the subapertures (see [3,7]).\\
The AO system works in a closed loop at 500 Hz and there is two-step delay between measurement and correction. We assume that the DM has an instantaneous response and the coupling factor of the actuators is 0.3. With a zonal basis, the phase is estimated only on the actuators' locations: the number of valid actuators is 241 for the 8 m telescope and 877 for the 16 m telescope.\\
Phase screens are generated respectively on a 160$\times$160 or on a 320$\times$320 grid, with 10 times 10 points per each sub aperture: the correction phase is therefore obtained by multiplying the prediction estimate on the actuators' locations with the influence matrix.\\
For the turbulence temporal model in the KF, in the ETKF and in the local ETKF-based control laws, we have chosen a \textit{first} order Auto Regressive model (AR1) [1-2].\\
Each value of the coherent energy has been calculated with several simulations of 5000 iterations (10 sec).

\subsection{The ETKF compared with the KF-based optimal control law (D = 8 m)}

This simulation (see figure 2) was done with three different values of noise variance: in each case, the curve gives the loss of performance in term of coherent energy between the ETKF-based control law and the optimal solution given by the KF. This loss is depending on the number of members. This simulation result confirms the theoretical result: when the number of members increases, in term of coherent energy, the performance provided by the ETKF converges to the one provided by the KF optimal solution whatever is the value of the noise variance.
\begin{figure}[!ht]
\centerline{\resizebox{0.95\columnwidth}{!}{\rotatebox{270}{\includegraphics{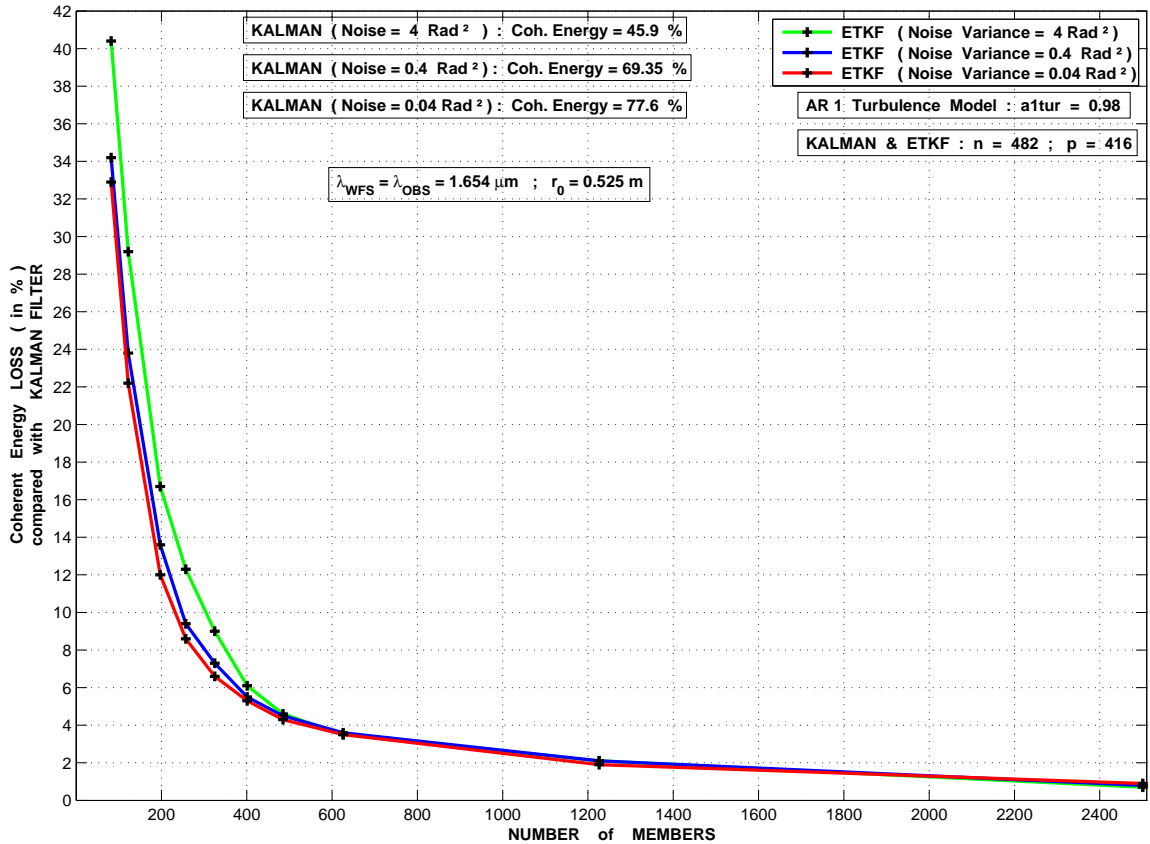}}}}
\caption{Loss of coherent energy with the ETKF compared with the KF (D = 8 m)}
\label{fig:2}
\end{figure}

\subsection{The ETKF and the local ETKF compared with the KF-based optimal control law (D = 16 m)}

This simulation (see figure 3) was done with one value of noise variance: in each case, the curve gives the loss of performance in term of coherent energy between one control law (ETKF, local ETKF \textit{without} removing or \textit{by} removing differential pistons) and the optimal solution given by the KF. This loss is depending again on the number of members.\\
This simulation result shows that:\\
- in term of coherent energy for a given number of members, the performance provided by the \textit{local} ETKF is better than the one provided by the ETKF.\\
- with the \textit{local} ETKF, in the phase estimate, there is still a problem of reattachment between the various domains: the update estimates of the turbulent phase are done separately on \textit{each} domain. Therefore, we have to reconstruct the entire phase estimate on the whole pupil with all these various small phase estimates and it appears differential pistons between all various domains. We have implemented a first \textit{post-processing} method which removes these differential pistons. This simulation result shows that, for a given number of members, the performance of the local ETKF with this basic method is improved and we can achieve a loss of coherent energy compared to KF less than 3 \% with 197 members.
\begin{figure}[!ht]
\centerline{\resizebox{0.95\columnwidth}{!}{\rotatebox{270}{\includegraphics{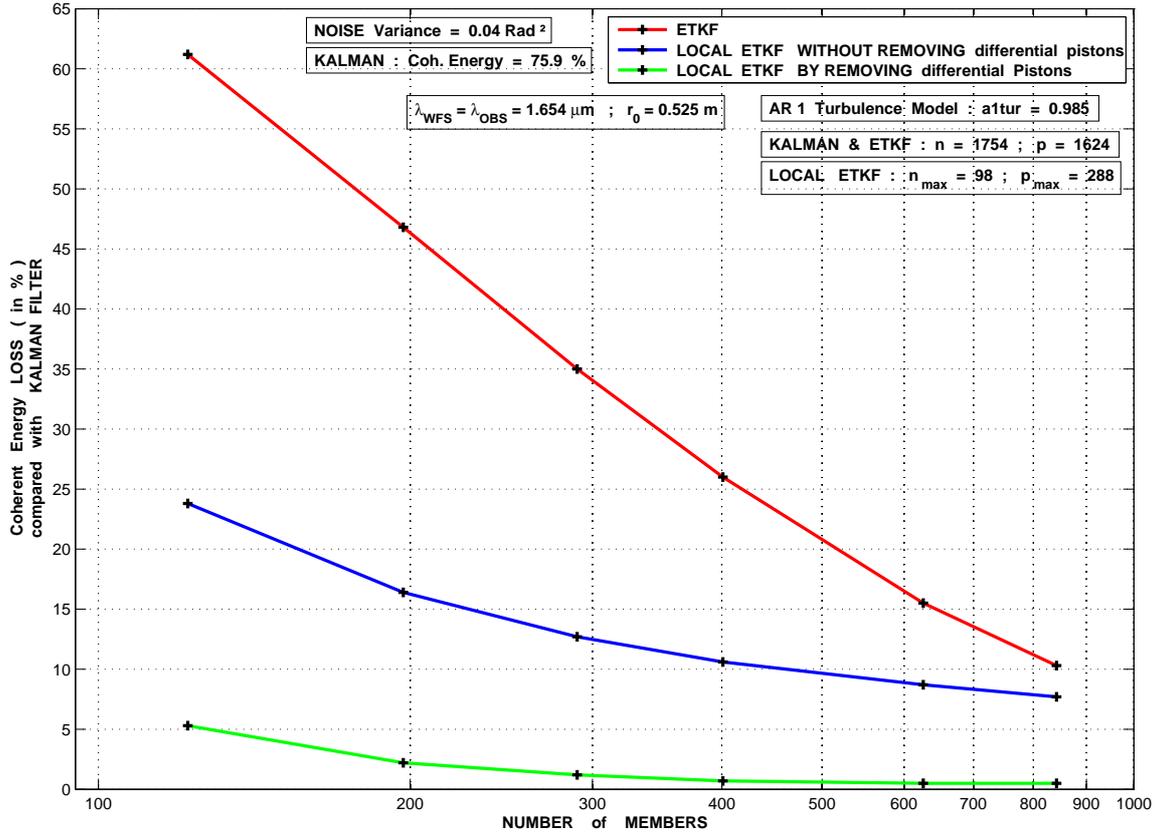}}}}
\caption{Loss of coherent energy with the ETKF \& the local ETKF compared with the KF (D = 16 m)}
\label{fig:3}
\end{figure}
\\
Moreover, in term of numerical complexity, there is a real gain with the local ETKF as this control law is implemented with a parallel architecture for which update step calculations are completely independent. We can illustrate with 2 examples:\\
- in the case of a 16 m telescope with this domain decomposition (see figure 1), all domains don't have the same number of actuators. Therefore, the dimensions of all estimate vectors on all estimation domains and the dimensions of all observation vectors on all observation regions are not the same. However, the largest value of all dimensions of the estimate vectors is $n_{max} = 98$, and the largest value of all dimensions of the observation vectors is $p_{max} = 288$. These values have to be compared with those used in the KF and the ETKF-based control laws: the dimension of the estimate vector on the whole pupil is $n = 1754$, and the dimension of the observation vector on the whole pupil is $p = 1624$. With the \textit{local} ETKF, the theoretical complexity is now given by $O(m^{3} + m^{2}\times(n_{max} + p_{max}))$: we can notice that, as $n_{max}$ and $p_{max}$ are smaller than the values of \textit{n} and \textit{p}, with respectively a reduction factor of 17 and 5, the numerical complexity will be significantly reduced.\\\
- in the case of a 40 m telescope with 69 domains, we have: $n = 10354$, $p = 10032$ with the KF \& the ETKF and $n_{max} = 162$, $p_{max} = 242$ with the local ETKF. The previous reduction factors become respectively 65 and 40. They can be much larger if we take more small domains: the numerical complexity can be therefore dramatically reduced. With the \textit{local} ETKF, on each of these 69 domains, there will be more than 40 times less multiplications than the total number of multiplications computed with the ETKF.\\
Of course, the real performances will depend on the relative magnitudes of $n_{max}$, $p_{max}$ and \textit{m} and the effective parallel implementation of this algorithm.


\section{Parallelization and preliminary speed tests with the local ETKF}\label{sec:3}

As mentioned before, the local ETKF is an intrinsic parallel algorithm and moreover, computations during the update step for each domain are matrix operations which can further be parallelized. The prediction step can also be easily parallelized.\\
We have already a parallel implementation on our cluster with CPUs, which has given the real number of floating point operations required with the local ETKF-based control law.\\
We can take the reference case with a 40 m diameter telescope, a 69 domains decomposition and 197 members in the ensemble. The update step will require a total of 5 GFLOP ($5.10^9$ floating point operations); in fact, during this update step, each domain will require only 0.08 GFLOP at maximum. The prediction step will require a total of 0.02 GFLOP.\\
This means that, in order to operate at a 500 Hz frequency, we will need 2.5 TFLOP/sec (or a maximum of 40 GFLOP/sec on each domain) for all update steps, and 10 GFLOP/sec for all prediction steps.\\
Two possible architectures can be considered:\\
- the first one is a computation cluster where each node makes all computations for each domain. In our reference example, each node should have a performance of 50 GFLOP/sec. During each cycle of the AO loop, the total amount of data which should be transmitted via network is about 10 Mbyte, which is about 5 Gbyte/sec (40 Gbit/sec) at a 500 Hz frequency. It is almost one order of magnitude less than the bandwidth of already existing solutions. Our preliminary tests show that, on a workstation with 4-cores processor (Intel Xeon W3565), calculations of the update step for one domain takes about 0.005 sec, which is only 2.5 times more than what will be required for a 500 Hz frequency. This architecture is therefore possible with nowadays computers.\\
- the second one is a single computer + GPUs. This solution can be much cheaper than computation cluster. Modern GPUs are claimed to have peak performance up to a few TFLOP/sec. However, we have not implemented the local ETKF on GPUs  yet (this is a work in progress) so that we can not give here the results on real speed tests. Nevertheless, we assume that our algorithm can be efficiently implemented on GPUs. We will give the performances on this type of architecture in a future simulation.


\section{Conclusions and perspectives}\label{sec:4}

As a short conclusion, we would like to recall that the local ETKF is a new algorithm for complex AO systems control law on ELTs. We have already implemented a parallel version on our cluster with CPUs. In this paper, we have shown that we can:\\
- keep the KF formalism which enables to reach the performances of an optimal control law,\\
- have a non-stationary model of the turbulence which enables to deal with the evolution of the turbulence during all the observation period,\\
- efficiently implement this control law on a distributed parallel environment (with CPUs) and nowadays computers performances are enough to use it with 40 m class telescopes with some particular domain decompositions,\\
- have linear numerical complexity on both dimensions $n_{dom}$ and $p_{dom}$ on each domain.\\
As all calculations on each domain can be performed separately, we tested a version with two parallel local ETKF running together with two different actuators domain partitions. This can really improve the performances: in particular, we are developing a new method which will remove the differential pistons inside the AO loop. Of course, we are also studying the possibility of implementing the local ETKF-based control law on GPUs architecture. 


\section*{Acknowledgements}

This work was supported by financial grants from the cross-disciplinary mission Mastodons (CNRS) and from the PHC Aurora (Programme Hubert Curien-Campus France).


\section*{References}

[1] C. Kulcsar et al., "Optimal control, observers and integrators in adaptive optics", Opt. Express \textbf{14}, 7464-7476 (2006).\newline
[2] B. Le Roux et al., "Optimal control law for classical and multiconjugate adaptive optics", JOSA A \textbf{21}, 1261-1276 (2004).\newline
[3]	C. Petit et al., "LQG control for adaptive optics and MCAO: experimental and numerical analysis", JOSA A \textbf{26}, 1307-1325 (2009).\newline
[4] A. Costille et al., "Wide field adaptive optics laboratory demonstration with closed-loop tomographic control", JOSA A \textbf{27}, 469-483 (2010).\newline
[5] M. Gray, B. Le Roux, "Ensemble Transform Kalman Filter: towards a dynamic and optimal control law for adaptive optics on ELTs", Proceedings \textbf{AO for ELT II} (2011).\newline
[6] M. Gray, B. Le Roux, "Ensemble Transform Kalman Filter, a non-stationary control law for complex adaptive optics systems on ELTs: theoretical aspects and first simulation results", Proceedings \textbf{SPIE 8447-65} (2012).\newline
[7] Massioni et al., "Fast computation of an optimal controller for large scale adaptive optics", JOSA A \textbf{28}, 2298-2309 (2011).\newline
[8] C. Bishop et al., "Adaptive sampling with the Ensemble Transform Kalman Filter. Part I: Theoretical Aspects", Monthly Weather Review \textbf{129}, 420-435 (2001).\newline
[9] P. Sakov, P. Oke, "Implications of the Form of the Ensemble Transformation in the Ensemble Square Root Filters", Monthly Weather Review \textbf{136}, 1042-1053 (2008).\newline
[10] J. Mandel, "Efficient implementation of the Ensemble Kalman Filter", University of Colorado, Center for Computational Mathematics Reports \textbf{231}, (2006).\newline
[11] P. Oke, "Impacts of localisation in the EnKF and EnOI: experiments with a small model", Ocean Dynamics \textbf{57}, 32-45 (2007).\newline
[12] J. Anderson, "Exploring the need for localization in ensemble data assimilation using a hierarchical ensemble filter", Physica D \textbf{230}, 99-111 (2007).\newline
[13] C. Bishop, D. Hodyss, "Flow-adaptive moderation of spurious ensemble correlations and its use in ensemble-based data assimilation", Q. J. R. Meteo. Soc. \textbf{133}, 2029-2044 (2007).\newline
[14] B. Hunt et al., "Efficient data assimilation for spatiotemporal chaos: a Local Ensemble Transform Kalman Filter", Physica D \textbf{230}, 112-126 (2007).\newline


\end{document}